\documentclass[11pt, a4paper, logo, copyright, nonumbering]{xiaomi}

\usepackage[authoryear, sort&compress, round]{natbib}
\usepackage{dblfloatfix}
\usepackage{ulem}
\usepackage{caption}
\usepackage{dramatist}
\usepackage{xspace}
\usepackage{pifont}
\usepackage{multirow}
\usepackage{tcolorbox}
\usepackage{xltabular}
\usepackage{longtable}
\usepackage{hyperref}
\usepackage{wrapfig}
\usepackage{graphicx}

\usepackage{amsfonts}
\usepackage{amsmath}
\usepackage{amssymb}
\usepackage{lineno}
\usepackage{multirow}
\usepackage{adjustbox}

\usepackage[bottom]{footmisc}

\usepackage{CJKutf8}
\usepackage{subcaption}
\usepackage{setspace}
\usepackage{makecell}
\usepackage{graphicx}
\usepackage{multicol}
\usepackage[defaultlines=2,all]{nowidow}

\defcitealias{nondeter}{T. M. Lab}

\definecolor{xiaomiorange}{HTML}{FF6901}

\title{Full-Pipeline Inference Optimization for MiMo-V2.5 Series: Pushing Hybrid SWA Efficiency to the Limit}
\begin{abstract}

We present a full-pipeline inference optimization for the MiMo-V2.5 model family, which combines Hybrid Sliding Window Attention (Hybrid SWA), sparse Mixture-of-Experts (MoE), and multimodal encoders. While Hybrid SWA can ideally reduce both attention compute and KVCache storage significantly compared to Full Attention, realizing these gains in production requires substantial engineering effort. We systematically optimize the KVCache system with layerwise prefetch, SWA-aware prefix cache trees, and specialized placement strategies, achieving strict $O(W)$ SWA storage and high cache hit rates. We further build GCache, a high-performance distributed cache infrastructure with RDMA-optimized networking, and develop a KVCache-affinity router to reduce computation while preserving load balancing. We also optimize for multimodal inputs, including GPU image preprocessing, parallel video decoding, and multimodal cache sharing.
Together, these optimizations constitute the first large-scale LLM serving system in production that efficiently covers the Hybrid SWA + MoE + multimodal composite architecture.

\end{abstract}

\begin{document}

{
    \bgroup
    \setlength{\parindent}{0pt}
    \vspace*{3pt} 
    \begin{adjustwidth}{0pt}{0pt}  
    \begin{center} 
    {\titlefont Full-Pipeline Inference Optimization for MiMo-V2.5 Series: Pushing Hybrid SWA Efficiency to the Limit \par}
    {
    \vskip5pt
    {\normalfont\sffamily\fontsize{11}{15}\selectfont MiMo Team, Xiaomi}
    \vskip10pt
    }
    \end{center}
    \end{adjustwidth}
    \egroup
    {\abscontent}
    \thispagestyle{firststyle} 
}

\renewcommand{\thefootnote}{\fnsymbol{footnote}}
\renewcommand{\thefootnote}{\arabic{footnote}}

\section{Introduction}

The MiMo-V2.5 model family, including MiMo-V2.5~\citep{mimov25} and MiMo-V2.5-Pro~\citep{mimov25pro}, combines several architectural design choices: Hybrid Sliding Window Attention (Hybrid SWA) compresses KVCache storage to roughly 1/7 that of Full Attention; sparse MoE activation cuts per-token compute while preserving model capacity; and multimodal encoders enable cross-modal understanding across vision, audio, and video. Together, these features give the MiMo-V2.5 series significant performance and efficiency potential in long-context and multimodal scenarios.

From the outset, our goal was clear: train a model that is both powerful and efficient for long-context reasoning. These two objectives are inherently in tension. Strong reasoning requires modeling long-range dependencies, which typically demands larger-scale attention computation and higher KVCache overhead. In traditional Full Attention architectures, both attention compute and KVCache storage grow rapidly with context length, making long-context training and inference prohibitively expensive. Hybrid SWA works by interleaving local Sliding Window Attention (SWA) with global Full Attention across layers: most layers compute attention only within a local window, while a small number of key layers retain a global view. In theory, this structure reduces attention complexity to near-linear while preserving the ability to model long-range dependencies.

However, theoretical architectural advantages do not automatically translate into production efficiency. Hybrid SWA introduces new complexity in managing KVCache hit rates, prefix matching, and maintaining dual-semantic consistency between Full Attention and SWA layers. Real engineering systems face further challenges — data movement across multi-level storage, misaligned async prefetch and scheduling, difficulty synchronizing distributed cache states — that prevent theoretical gains from being directly achieved.

Beyond Hybrid SWA, MoE imposes significant demands on distributed scheduling and load balancing, while the multimodal encoders remain a throughput bottleneck in large-image and long-video scenarios. Scheduling strategy and the Prefill/Decode execution pipeline also require careful optimization. This article presents an end-to-end engineering practice for the inference system of the MiMo-V2.5 series, covering KVCache management, tiered caching systems, SWA-aware prefix cache trees, scheduling strategies, Prefill/Decode execution pipelines, and multimodal optimizations — systematically realizing the architecture's theoretical efficiency potential (especially Hybrid SWA) in production.
\section{Background}

Before diving into specific optimizations, let's first quantify the theoretical efficiency bounds of Hybrid SWA — the architectural rationale behind the design choice and the baseline against which all subsequent optimizations are measured.

\subsection{Compute Analysis}
Taking MiMo-V2.5-Pro as an example, the model has 70 layers in total: 10 Full Attention layers and 60 SWA layers, with a sliding window size of 128. Compared to Full Attention, the compute cost of Hybrid SWA is illustrated in the figure below. SWA layers account for 6/7 of all layers, so the total compute of the Hybrid SWA architecture is roughly 1/7 that of Full Attention. In Chunked Prefill scenarios, where prefill is largely compute-bound, this directly translates to a proportional reduction in prefill cost.

\subsection{KVCache Storage Analysis}
Since SWA layers only need to retain KV within the sliding window — not for the full sequence — KVCache memory usage similarly drops close to 1/7. The decode phase is predominantly memory-bound, and its latency is proportional to the combined bytes read for model parameters and KVCache. For long sequences, KVCache volume can far exceed model parameters, so the reduction in KVCache storage translates almost directly into a reduction in decode cost in long-sequence scenarios (except for models with sparse attention, which reduces per-token KV access).

\begin{figure}[t]
\centering
\begin{subfigure}[t]{0.45\textwidth}
    \centering
    \includegraphics[width=\textwidth]{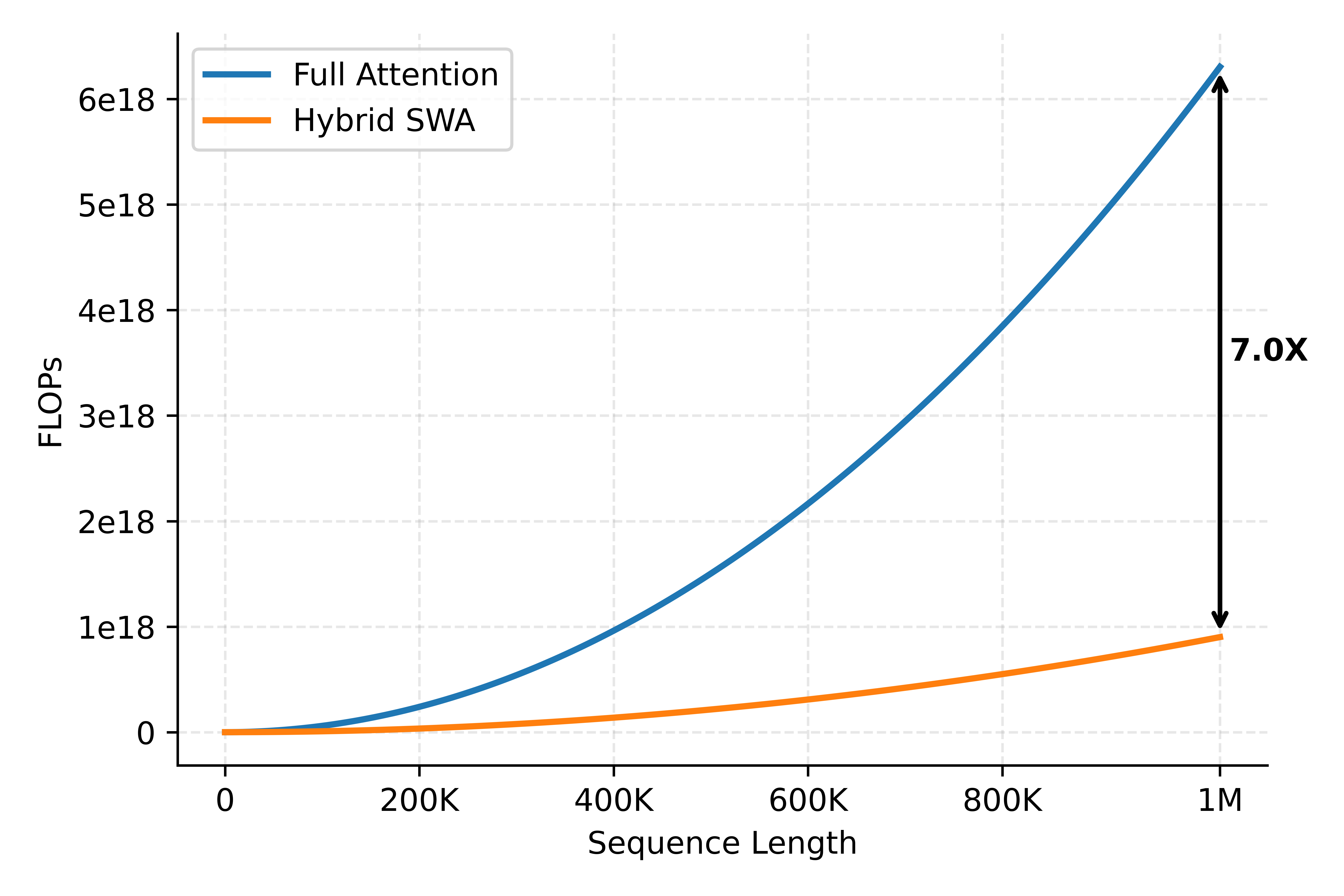}
    \caption{Attention FLOPs vs.\ sequence length.}
\end{subfigure}
\begin{subfigure}[t]{0.45\textwidth}
    \centering
    \includegraphics[width=\textwidth]{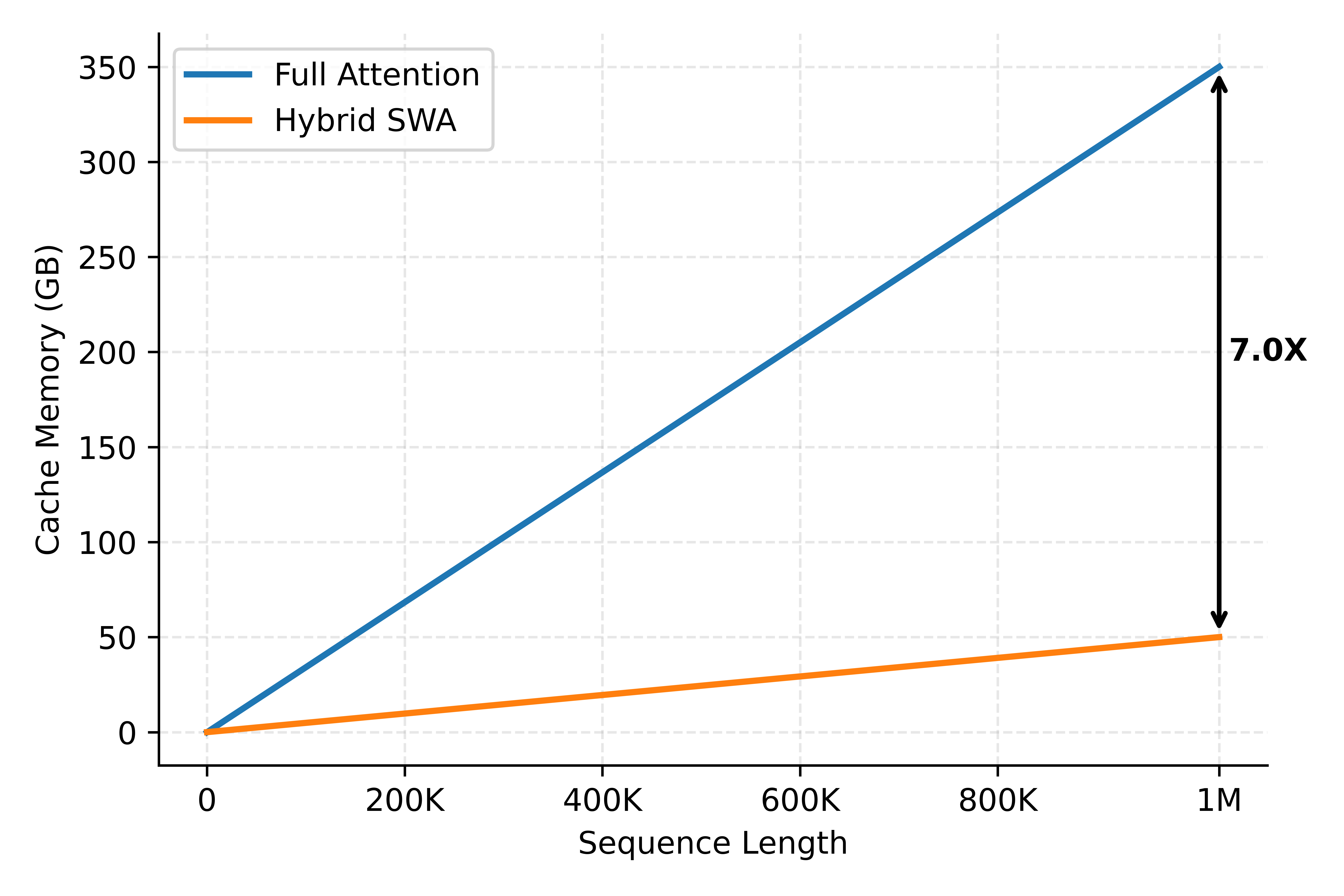}
    \caption{KVCache memory vs.\ sequence length.}
\end{subfigure}
\caption{Theoretical efficiency analysis of Hybrid SWA on MiMo-V2.5-Pro. Both attention compute and KVCache storage are reduced by approximately $7\times$ compared to Full Attention.}
\label{fig:analy_2}
\end{figure}

\begin{figure}[t]
\centering
\begin{subfigure}[t]{0.45\textwidth}
    \centering
    \includegraphics[width=\textwidth]{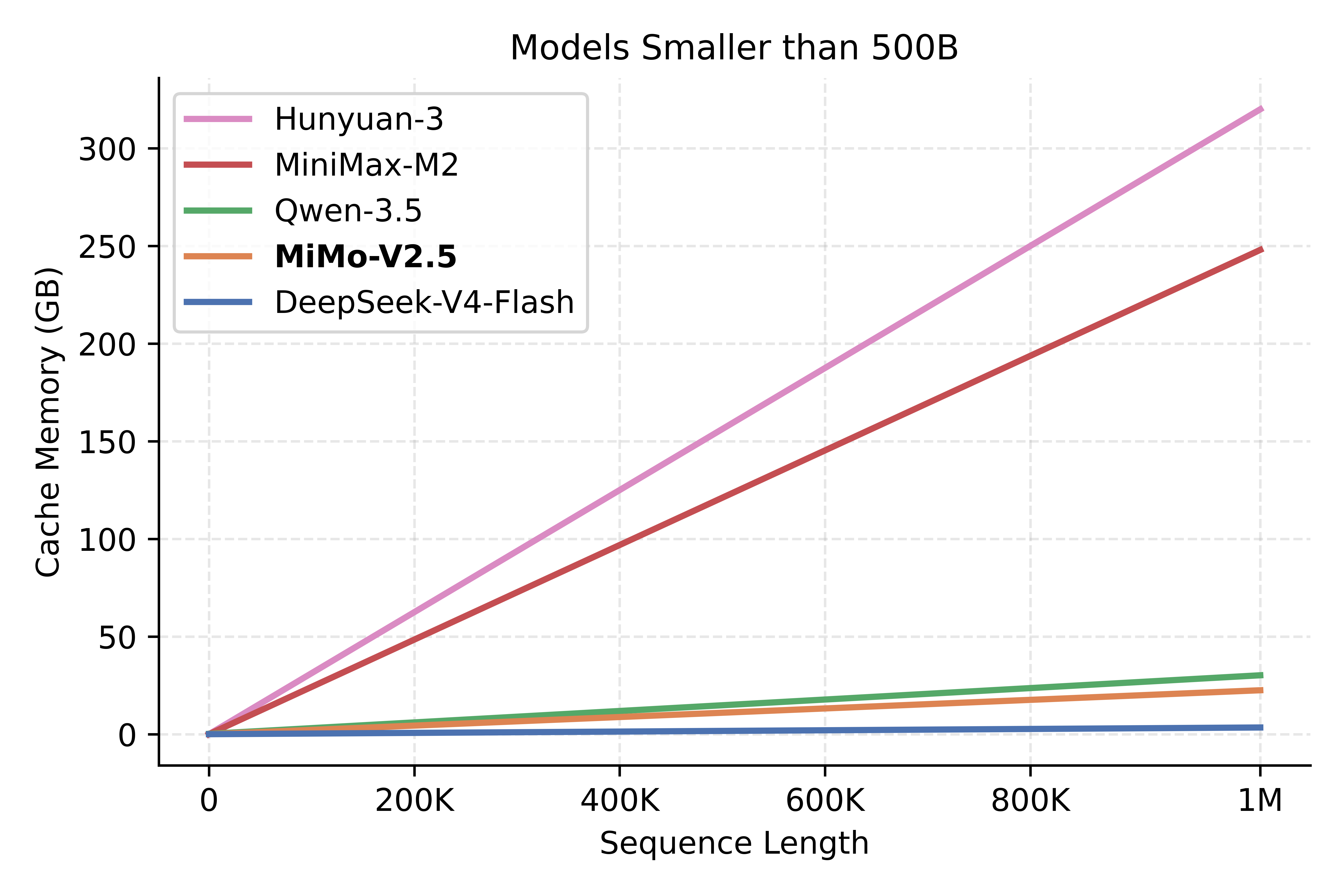}
    \caption{Models under 500B parameters.}
\end{subfigure}
\begin{subfigure}[t]{0.45\textwidth}
    \centering
    \includegraphics[width=\textwidth]{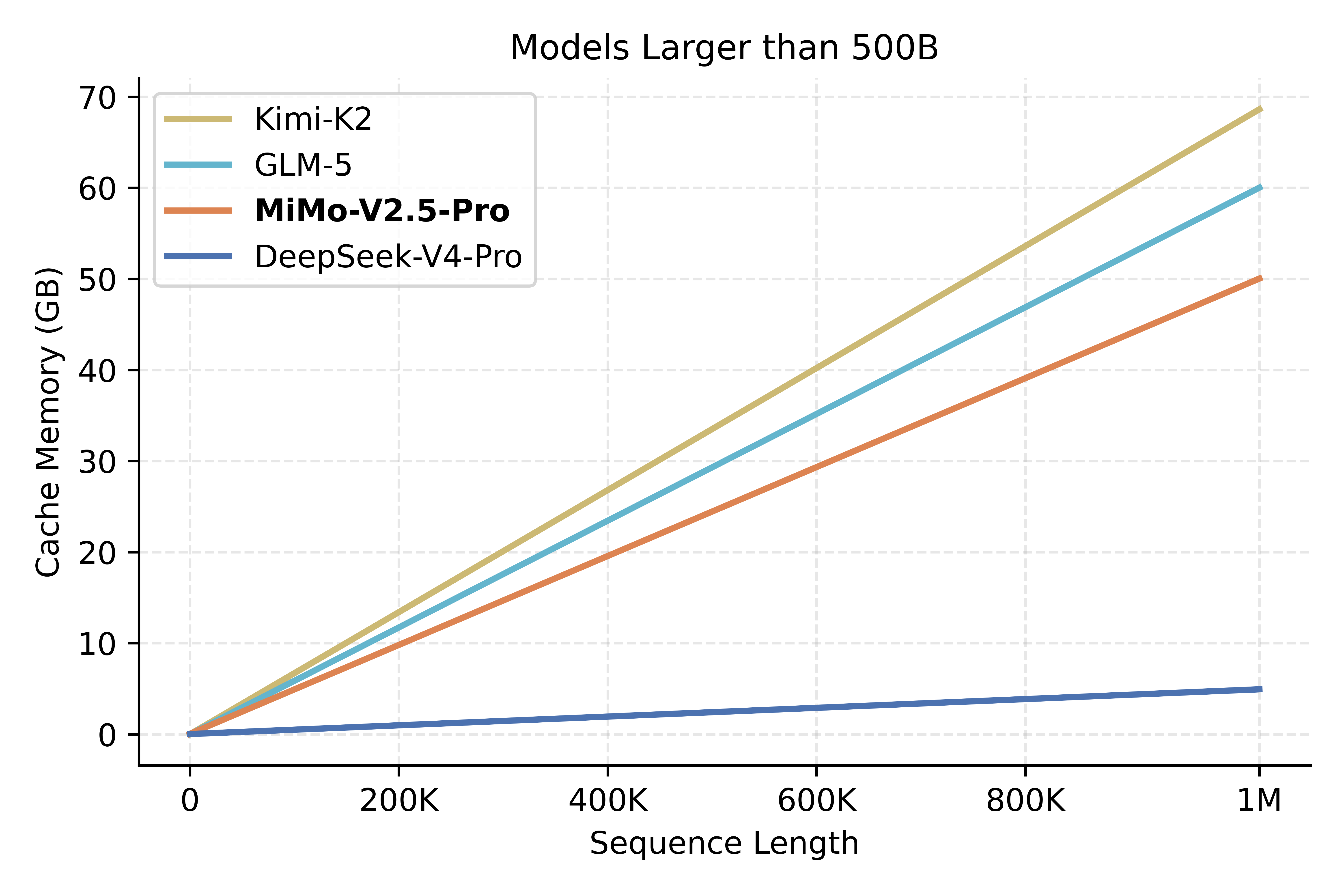}
    \caption{Models over 500B parameters.}
\end{subfigure}
\caption{Estimated KVCache memory vs.\ sequence length across different model architectures.}
\label{fig:kvcache_comparison}
\end{figure}

KVCache storage varies substantially across model architectures. Figure~\ref{fig:kvcache_comparison} compares representative models in two parameter-scale groups: models below 500B parameters and models above 500B parameters. The model configurations are obtained from their official checkpoints~\citep{mimov25pro,mimov25,deepseekv4pro,deepseekv4flash,glm5,kimik2,qwen35,minimaxm2,hy3}. Within their respective groups, MiMo-V2.5 and MiMo-V2.5-Pro have the second-lowest estimated KV cache memory requirements, behind only DeepSeek-V4-Flash and DeepSeek-V4-Pro, respectively.

It is worth noting that actual cost differences do not strictly correspond to KVCache size ratios, as there are fixed compute and memory access costs independent of sequence length. However, in long-context scenarios, the overall trend holds: \textbf{the gains are marginal for short sequences, but the longer the sequence, the greater the inference cost advantage}.
\section{KVCache System Refactor}

The MiMo-V2 and MiMo-V2.5 series were among the earliest models to adopt the Hybrid SWA architecture, but at the time, neither mainstream open-source inference frameworks nor caching systems offered complete SWA support. 
When we launched the MiMo API, we chose SGLang v0.5.5~\citep{sglang} as the serving backend codebase --- and immediately encountered a severe challenge. 
In that version, SGLang's HiCache did not support SWA, or rather, early SWA support was implemented by storing the full KVCache to maintain compatibility. While there were some workarounds to make SWA more usable, we wanted to build a KVCache system with higher performance ceilings and better usability.

\subsection{SWA KVCache Management}

\subsubsection{KVCache Dual-Pool Design}

Hybrid SWA introduces a fundamental storage conflict: Full Attention layers require storing the full sequence KV ($O(N)$), while SWA layers only need to maintain KV within the sliding window ($O(W)$). Under a traditional single KV pool design, the system must allocate GPU memory at $O(N)$ for all layers, preventing the window sparsity of SWA from being leveraged --- effectively degenerating into a near-full KVCache implementation.

A natural solution is to split the KVCache into two independent pools for Full Attention and SWA, with unified abstraction at the system level:

\begin{itemize}
    \item \textbf{Physical layer:} Maintain separate Full KV pool and SWA KV pool. The SWA pool is sized only for the window and supports independent eviction based on the window, strictly constraining SWA storage to $O(W)$. This mechanism extends to L2 and L3 storage tiers as well.
    \item \textbf{Logical layer:} Expose a single sequence view to upper layers (prefix tree, scheduler, transport protocol), with the Full Attention index as the authoritative reference and a Full $\to$ SWA mapping maintained for transparent tiered storage.
    \item \textbf{Scheduling constraints:} The system validates both Full KV and SWA KV capacity constraints when admitting requests, avoiding resource misallocation from single-dimensional checks.
    \item \textbf{Data movement:} Cross-tier transfers are performed based solely on the SWA mask, ensuring only valid window data is moved and avoiding redundant bandwidth consumption.
\end{itemize}

Through this design, SWA KVCache achieves strict $O(W)$ storage constraints at the system level, improving overall KVCache capacity efficiency by \textbf{approximately $7\times$} and unlocking the structural advantages of Hybrid SWA. Mainstream inference frameworks have also adopted similar implementation approaches.

\subsubsection{Layerwise KVCache Prefetch}

With the SWA KVCache storage optimization in place, SWA layers only need to prefetch a minimal amount of KVCache. This enables near-perfect overlap between Host-to-Device KVCache prefetch and computation through layerwise scheduling, bringing the cost of cache reads during inference close to zero.

\begin{figure}[t]
\centering
\includegraphics[width=0.85\textwidth]{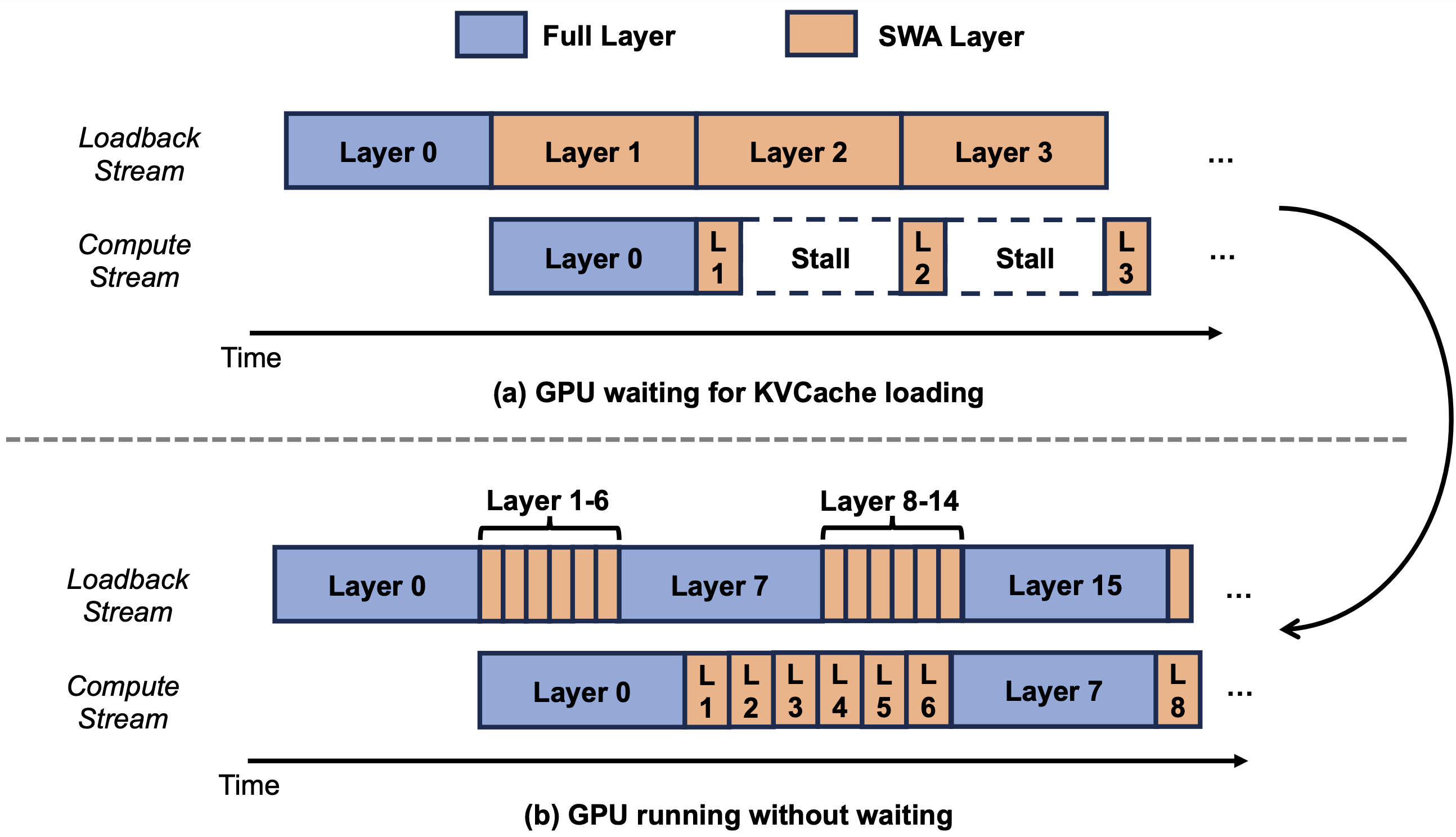}
\caption{Layerwise KVCache prefetch: (a)~the compute stream stalls waiting for KVCache loading; (b)~SWA-aware layerwise scheduling overlaps loadback and compute so GPU runs without waiting.}
\label{fig:layerwise_prefetch}
\end{figure}

\subsubsection{SWA-Aware Prefix Cache Tree}

The traditional RadixAttention hit rule is built on a simple assumption: equal token sequences $\to$ equal KV. This assumption holds under Full Attention --- as long as two requests share the same token IDs, their corresponding KV is guaranteed to still be in the pool and directly reusable.

But this assumption breaks under SWA. The reason is that the logical lifecycle of the prefix tree and the physical lifecycle of SWA KV are misaligned. Prefix tree node lengths are not constrained by the SWA window --- a node's sequence length can be shorter than the window or far longer, and nodes change continuously through request merging, splitting, and removal. As a result, a prefix tree node may still logically represent a complete token sequence, but its corresponding SWA KV may have only the tail portion remaining, or may have been evicted entirely. If the prefix tree still provides reuse length based on the ``token equality $\to$ hit'' rule, the scheduler may receive a pseudo-hit with evicted tail KV --- subsequent attention computation would read invalid or overwritten slots, directly degrading model correctness.

To keep prefix reuse correct and efficient under SWA, the prefix tree semantics must be revised in three ways:

\begin{enumerate}
    \item \textbf{Matching rules upgraded to ``window-safe length'':} In addition to token equality, the tail $W$ tokens must still have valid slots in the SWA pool. The match length is clipped to this new boundary --- anything beyond it is treated as a miss. This ensures that KV retrieved from a hit segment is always valid.
    \item \textbf{Eviction tied to request lifecycle:} Completion of each chunk in long prefill, request termination, and every $N$ generated tokens during decode all trigger an out-of-window SWA release. This keeps SWA pool usage constant at $W$ or chunk-level magnitude during long-context/long-output tasks, rather than growing with sequence length.
    \item \textbf{Nodes carry dual indices:} Each prefix tree node records two sets of information --- the Full Attention segment index (determining logical order, participating in Full Attention layer computation) and the SWA segment mapping (determining window safety). Eviction is managed separately: window-outside SWA segments can be evicted independently while preserving Full Attention segments (keeping the prefix reusable by Full Attention layers), or the entire segment can be evicted.
\end{enumerate}

SWA's compression of KV volume to 1/7 is a capacity-level benefit, while hit rate is a reuse-level benefit. Together, they determine the actual prefill compute cost curve. After introducing the ``window-safe length'' matching rule, the raw hit rate for a given token capacity decreases slightly --- but the number of tokens that fit within the same storage budget grows several-fold. \textbf{Measured against a fixed storage budget, the effective hit rate improves dramatically.}

\begin{figure}[t]
\centering
\includegraphics[width=0.85\textwidth]{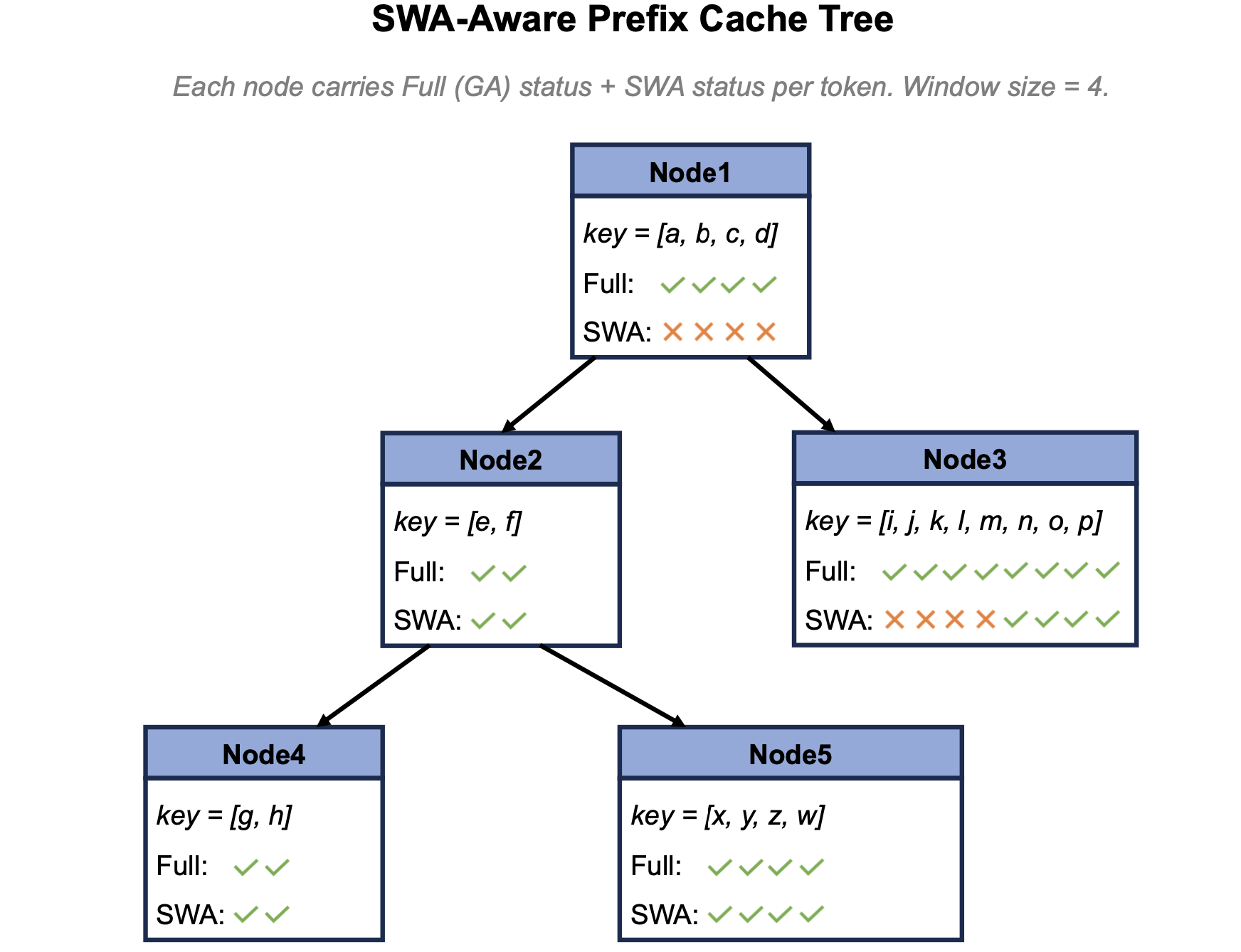}
\caption{SWA-aware prefix cache tree: each node carries per-token Full Attention status and SWA status, with window size 4; nodes track which tail tokens still have valid SWA slots.}
\label{fig:swa_prefix_cache}
\end{figure}

\subsubsection{KVCache Hit Rate Optimization}

After all three HiCache tiers are refactored to be SWA-aware, the device, host, and storage backend each maintain their own state of ``which positions have valid SWA.'' However, HiCache's data movement pipeline is asynchronous, caches across deployments differ, and shared prefix lengths across sessions also vary; the Full Attention Cache and valid SWA indices across tiers can easily fall out of sync. According to the SWA-aware prefix cache tree matching rules, if a sequence hits on the Full Attention Cache but misses on the SWA Cache, severe match-length truncation occurs: the more truncation, the longer the recomputation needed, and the lower the SWA Cache optimization effectiveness. We therefore optimized distributed consistency and cache hit rates across different scenarios:

\paragraph{Device complete, Host deficient.} When L3$\to$L2 prefetch only pulls in the tail segment due to bandwidth-latency tradeoffs, or when L1 prefix tree reorganization is not synced to L2/L3, this scenario arises. We proactively check the delta in SWA occupancy between device and host at timing points such as prefix tree node merging and prefill completion, allocate supplementary slots in the host's SWA pool, and asynchronously write device SWA KV via D2H transfer.

\paragraph{Host complete, Device deficient.} Naturally aligns at the next H2D transfer --- no active repair needed.

\paragraph{High-frequency sequence L3 prefix eviction.} Long sequence heads persist in L1/L2 due to high-frequency access, and cache affinity routes same-prefix requests to the same node. The L3 cache, due to long periods without direct access, may be evicted by the storage eviction policy --- prematurely releasing L3 Cache for globally high-frequency sequences and severely degrading cross-machine reuse. We periodically query L3 Cache when accessing L1/L2 Cache to prevent premature eviction.

\paragraph{Medium/short sequence SWA retention strategy.} Based on user request patterns, we retain relatively dense SWA KV Cache at fixed length positions for medium/short sequences. Although increasing SWA density raises the SWA ratio in overall KVCache, it directly benefits scenarios like multi-user shared system prompts.

Through these optimizations, we convert KVCache capacity expansion into longer effective hit lengths, making cross-session long-prefix reuse possible --- particularly beneficial for long agent sessions, multi-user shared system prompts, and repeated tool calls to the same codebase.

\subsection{GCache: High-Performance Distributed Cache Infrastructure}

GCache is a high-performance general-purpose cache system developed by the Xiaomi storage team, forming a critical part of unified training-inference storage architecture. Early on, during training scenarios, the storage team recognized that certain open-source caching projects provided limited acceleration for distributed file systems and could not fully exploit performance potential, so they began developing an in-house solution. Later, with the release of the MiMo large model and the launch of inference services, the team adapted GCache into an independent storage product for model distribution and as the L3 KVCache for the inference engine.

GCache supports both file and KV semantics, multi-level caching across memory/disk/remote tiers, shared-memory persistence and full-path zero-copy, high-concurrency non-blocking IO and RDMA communication, meeting upper-layer services requirements for high throughput and low latency while maintaining excellent scalability.

\subsubsection{Architecture Design}

\begin{figure}[t]
\centering
\includegraphics[width=0.85\textwidth]{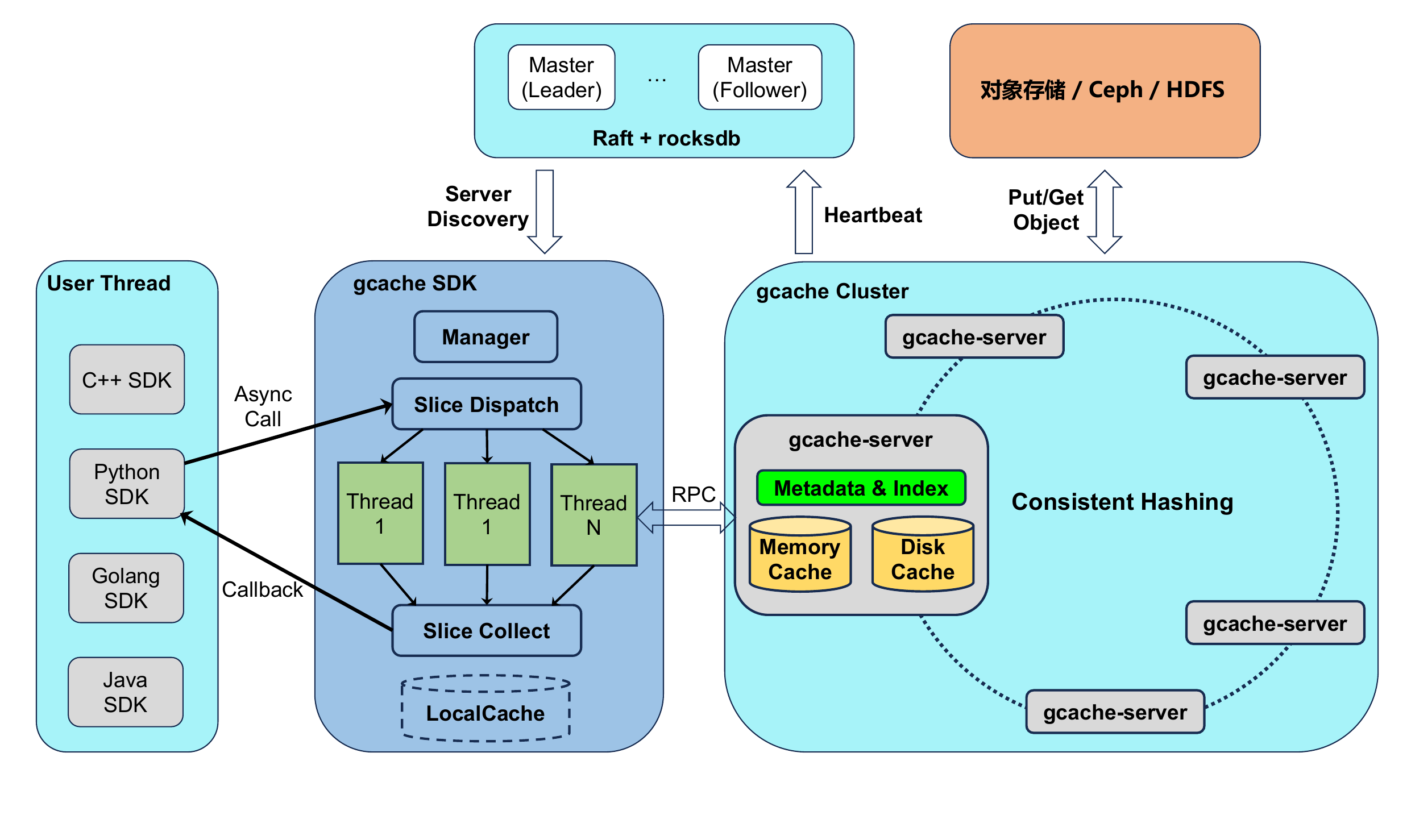}
\caption{GCache architecture: the SDK dispatches sliced requests to a cluster of gcache-servers organized by consistent hashing, with a Raft-based Master for service discovery and object storage (Ceph/HDFS) as the backend.}
\label{fig:gcache_arch}
\end{figure}

The overall architecture of GCache is shown in Figure~\ref{fig:gcache_arch}. GCache has several key features:

\begin{enumerate}
    \item \textbf{Decentralized metadata management} enables unlimited cluster scaling: Consistent hashing on keys determines storage locations. The Master uses a Raft-based highly-available deployment, but only manages heartbeats and service discovery --- IO paths do not pass through the Master.
    \item \textbf{Server-side support for both memory and disk caching:} Cold data in memory is evicted to disk; hot data on disk is promoted to memory. This approach is highly favorable for inference scenarios, automatically guaranteeing active session performance while reducing costs for long-idle sessions. Cache entries persist to shared memory --- no cache loss on service restart. Supports smooth scale-up or scale-down without cache loss.
    \item \textbf{Multi-language SDK with dedicated threads} for request slicing and dispatch: These threads do not consume user thread resources; slicing improves concurrency and keeps IO sizes within RDMA-friendly ranges. Threads use async callbacks with flexible callback granularity --- single KV level, batch level, or CUDA stream level.
\end{enumerate}

\subsubsection{Network Optimization}

Current mainstream GPU machines are equipped with $8\times$ 400G high-performance NICs. However, even with Prefill-Decode (PD)-disaggregated deployment, current inference frameworks struggle to saturate network bandwidth --- to the point where the industry is calling for reduced NIC specifications to cut costs.

To fully exploit high-speed networking, GCache prioritizes GPU NICs over frontend NICs for communication and performs extensive optimizations in the communication module, including NUMA binding and same-rail affinity. In benchmarks, with 1MB IO sizes, single-process RDMA read throughput reaches 170~GB/s at only 280~$\mu$s latency; under GDR scenarios, due to higher HBM bandwidth, single-process throughput reaches approximately 350~GB/s --- more than sufficient for inference framework communication requirements.

\subsubsection{Storage Cost Optimization}

2026 has seen growing industry concern about storage costs. Unlike other vendors using dedicated storage machines, GCache prioritizes co-deployment on GPU machines, taking over a portion of the memory from Prefill and Decode nodes along with the machines' built-in NVMe SSDs --- achieving zero additional storage cost.

\subsubsection{Reliability Assurance}

Due to co-deployment, the high failure rate of GPU machines poses a reliability challenge. Since launch, GCache has experienced host machine failures nearly every day. First, the team expended substantial effort hardening fault-handling logic. Second, since keys are fully distributed via consistent hashing, pre-grouping session IDs into logical sets ensures related sessions are spread across different nodes, reducing the blast radius of any single-node failure. Third, leveraging hardware detection capabilities from the underlying platform enables proactive fault discovery and automated data migration. For the rare sudden crashes that cannot be handled proactively, a short SDK timeout allows the inference framework to promptly detect misses and recompute, keeping online inference largely unaffected.

Based on these efforts, GCache maintains single-replica storage under co-deployment, without needing multi-replica redundancy for availability --- a key factor in its low storage cost.

\subsection{Discussion on Cache Hit Rate}

Thanks to the SWA KVCache optimizations described above --- lower storage footprint combined with a more stable, large-capacity GCache as L3 storage --- we were able to significantly extend Cache TTL (Time-To-Live) and improve KV Cache hit rates. KVCache eviction fundamentally stems from storage capacity constraints. As capacity nears saturation, the system prioritizes retaining KV Cache from new requests and evicts previously-accessed entries using LRU-like policies --- directly causing a given context to often miss when reused hours later. SWA's minimal storage footprint enables the same cost to hold several times more concurrent request caches, while large-capacity L3 further expands available capacity at low cost. The more storage space available, the less pressure on KVCache eviction, and the longer the retention duration. Longer TTL widens the hit window for historical contexts, and cache hit rates rise accordingly. Additionally, SWA's reduced bandwidth transfer overhead, while not directly affecting TTL, significantly lowers cross-tier data movement costs, ensuring stable and efficient operation of the entire caching system.

Since model launch, we have continuously observed on the server side: under mainstream high-quality harness frameworks, \textbf{server-side KV Cache hit rates average 93\%}; for heavy users with sustained high-intensity usage, this metric climbs even higher, reaching 95\% or above. Going forward, we will continue iterating SWA's KV Cache management logic and collaborate with more harness frameworks on harness-inference co-design to further optimize the hit rate ceiling.

\section{Scheduling Optimization}

In its early stages, the SGLang community's router service was not yet fully mature, with no shared state across instances. If a router service failed unexpectedly or requests were routed to a different router instance, KVCache scheduling would degrade. To solve this problem and ensure high availability in large-scale cluster deployments, Xiaomi developed LLM-Router --- a dynamically scalable stateless scheduler using Redis as centralized storage, eliminating KVCache degradation after single-service failures and consistently guaranteeing cache hit rates.

\subsection{KVCache and Load-Affinity Scheduling}

HiCache is highly sensitive to L2 hit rates. When L2 cache misses, the system must look up and fetch KVCache from L3, waiting for the fetch to complete before inference can begin. Improving L2 hit rates on the router side reduces unnecessary synchronous waits, directly boosting throughput.

The router implements KVCache affinity scheduling by maintaining dispatched requests in a Radix prefix tree. Among multiple Prefill instances, it prioritizes nodes that have already cached the current request's prefix while simultaneously balancing load to avoid load skew toward hotspots. After deployment, this strategy improved L2 cache hit rates by approximately \textbf{25\%} and per-node input throughput by approximately \textbf{30\%}. The core formula is roughly as follows:

\begin{equation}
\text{score}(\text{worker}) = \text{matchWeight} \times \text{prefixMatchPercentage} - \text{normalizedLoad}
\label{eq:router_score}
\end{equation}

\subsection{TTFT Optimization}

When model services experience queuing, the traditional FCFS (First Come First Serve) strategy does not consider the priority relationship between requests with higher and lower cache hit rates. Requests that have a higher cache hit rate but require less computation may end up waiting for lower-hit-rate requests to finish inference, causing TTFT P99 to become abnormally long and dragging down average throughput.

\begin{figure}[t]
\centering
\includegraphics[width=0.7\textwidth]{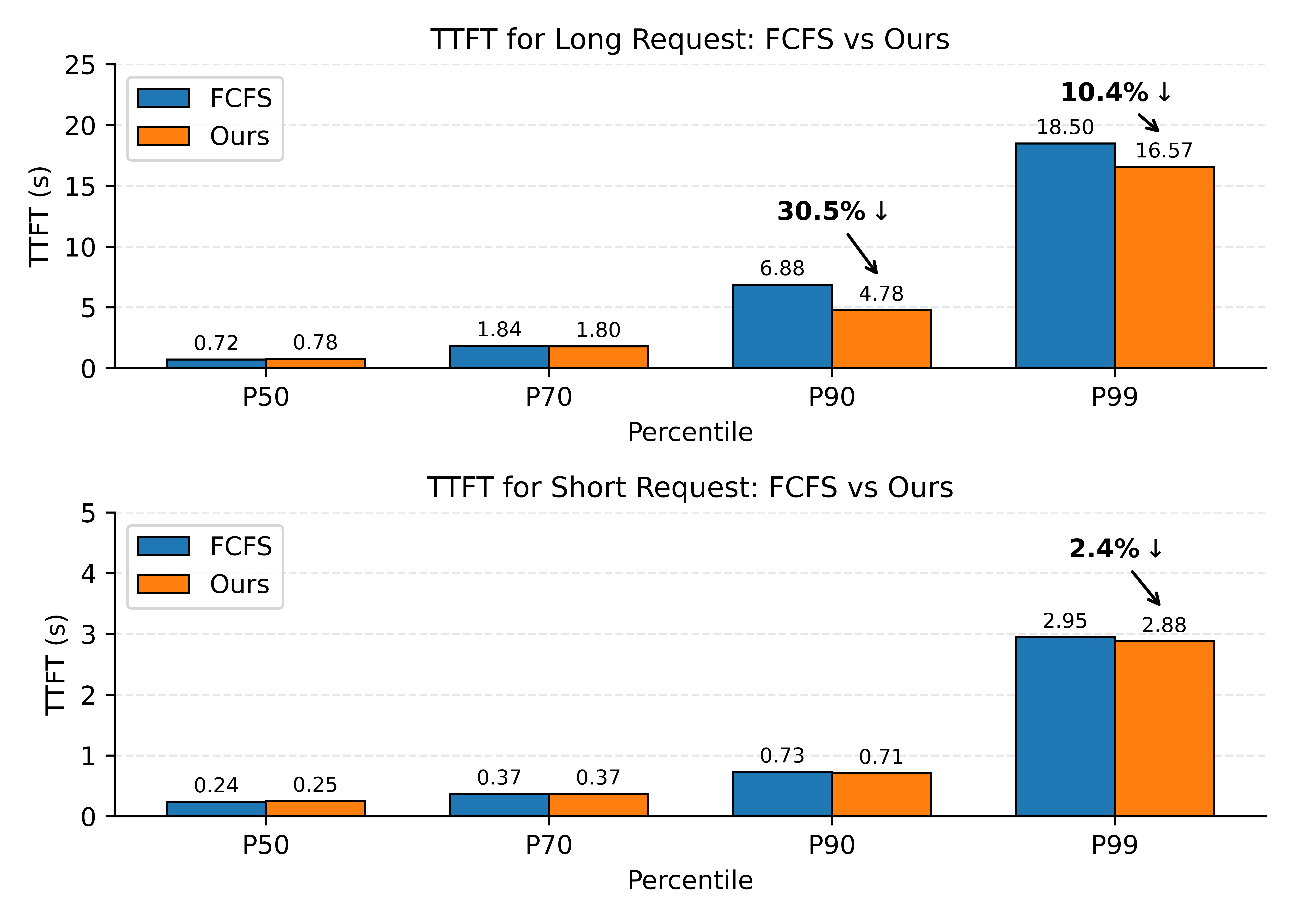}
\caption{TTFT comparison of FCFS vs.\ our scheduling strategy across P50/P70/P90/P99 for long requests (top) and short requests (bottom): long-request P90 drops 30.5\%, while short-request TTFT is essentially unchanged.}
\label{fig:ttft_compare}
\end{figure}

To address this, the router gives priority to requests with fewer uncached tokens when scheduling from the waiting queue, preventing cache-friendly requests from being blocked by slower ones and the resulting P99 degradation. However, this strategy can lead to starvation of certain requests, so we added a wait-time penalty mechanism to mitigate starvation. As shown in Figure~\ref{fig:ttft_compare}, our results show that this strategy does not degrade service quality for shorter requests, while \textbf{reducing TTFT P90 by up to 30\%} for longer ones.

\section{Prefill Optimization}

\subsection{Parallelism Configuration}

In theory, a smaller EP (Expert Parallelism) during the prefill stage yields better performance and throughput, in three ways: smaller cross-machine footprint and lower communication overhead; fewer DP (Data Parallelism) instances, reducing the impact of attention load imbalance between DPs; and more experts per machine, improving MoE load balance. However, EP size is constrained by GPU memory, which must accommodate both model parameters and KVCache. Previously, the SWA KVCache required storing KVCache for all tokens, forcing EP to be larger; after optimization, only tokens within the SWA window need to be stored, allowing us to reduce EP to half its original size, \textbf{improving end-to-end performance by approximately 40\%}. Going forward, we will continue exploring PP (Pipeline Parallelism) optimizations for the Hybrid SWA structure to further reduce EP size and improve overall throughput.

\subsection{Length Bucketing Strategy}

\begin{figure}[t]
\centering
\includegraphics[width=0.7\textwidth]{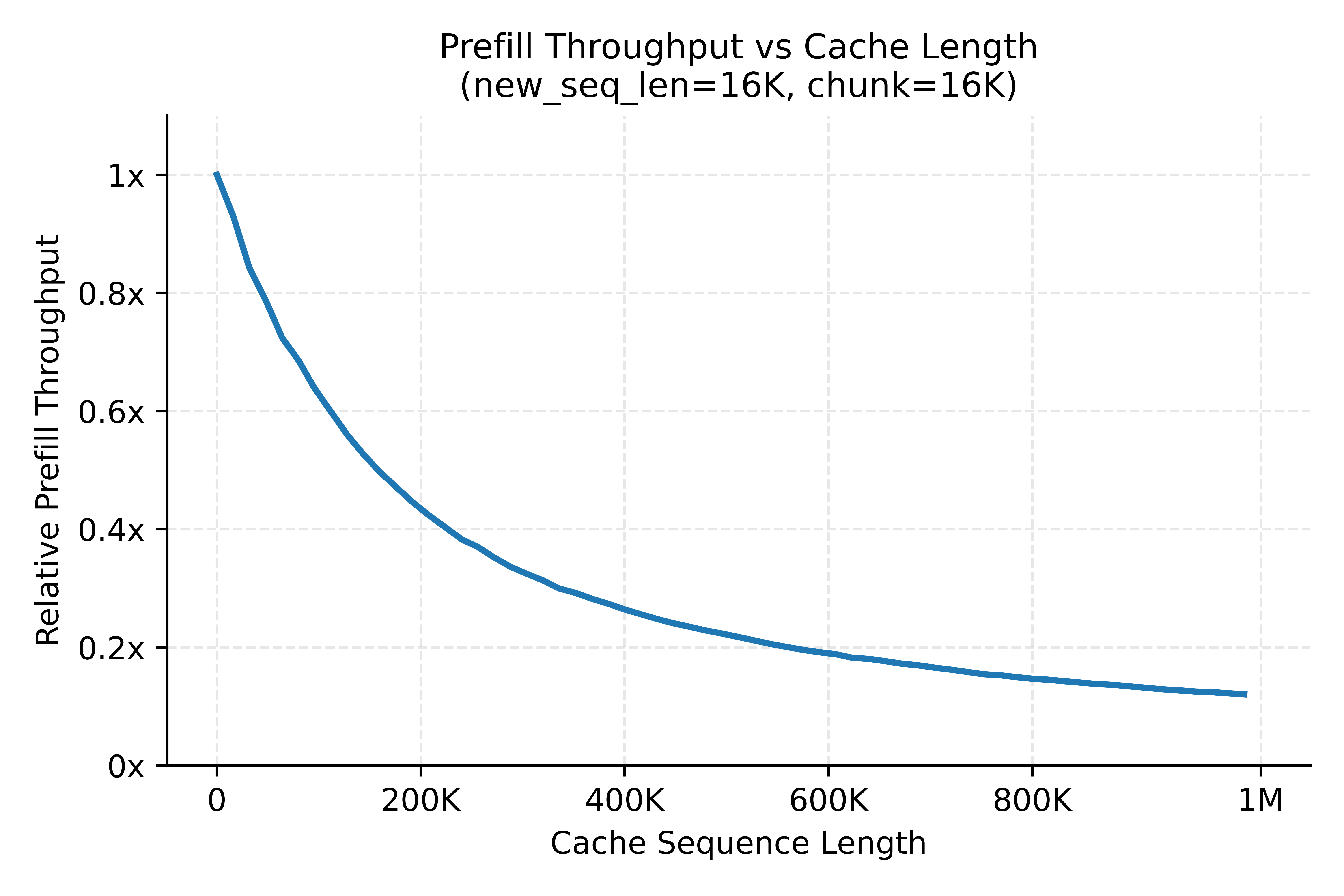}
\caption{Relative prefill throughput vs.\ cache sequence length with a fixed 16K compute chunk: throughput falls from $1\times$ near zero prefix to about $0.12\times$ at a 1M-token prefix.}
\label{fig:prefill_throughput}
\end{figure}

The MiMo-V2.5 series' hybrid architecture significantly improves compute efficiency over pure GQA, but throughput still degrades noticeably as sequence length increases. Figure~\ref{fig:prefill_throughput} shows throughput in Chunked Prefill with a fixed 16K-token compute chunk and prefixes of varying lengths.

In agentic scenarios, ultra-long requests mostly originate from multi-turn agent interactions with substantial prefix caches. When requests with significantly different lengths are scheduled to the same model instance, short requests are bottlenecked by long ones, degrading overall throughput in two main scenarios:

\begin{enumerate}
    \item \textbf{DP-Attention synchronization:} After each layer's attention computation, multiple DPs must synchronize via collective communication before entering the MoE stage. If long and short requests coexist across DPs in the same EP group, short requests are slowed by long requests' computation.
    \item \textbf{Chunked Prefill interference:} When requests with different prefix lengths are batched into the same chunk, short-prefix requests are dragged down by long-prefix requests' computation.
\end{enumerate}

To mitigate these load imbalance issues, we adopted a \textbf{three-tier length bucketing strategy} (0--64K / 64K--256K / 256K--1M), aggregating requests with similar load characteristics into the same bucket for computation, significantly improving average production prefill throughput. Building on this, we are currently exploring finer-grained, more flexible bucketing mechanisms to adapt to dynamic production workloads.

\subsection{MoE Load Balancing}

All MiMo-V2.5 series models use the MoE architecture, requiring consideration of expert load balancing during the prefill stage. Since the pre-training phase introduced load-balancing training objectives and the training process was relatively stable, the model learned a fairly uniform expert routing strategy. During inference, without enabling any expert load balancing strategy, the average expert load factor per layer (ratio of average token count across all ranks to the maximum token count of any rank in that layer) is approximately 0.85, already indicating a well-balanced distribution. Therefore, we currently do not incorporate any expert load balancing strategy. We will continue monitoring this metric and introduce related optimizations as needed based on evolving production load patterns.

\begin{figure}[t]
\centering
\includegraphics[width=0.7\textwidth]{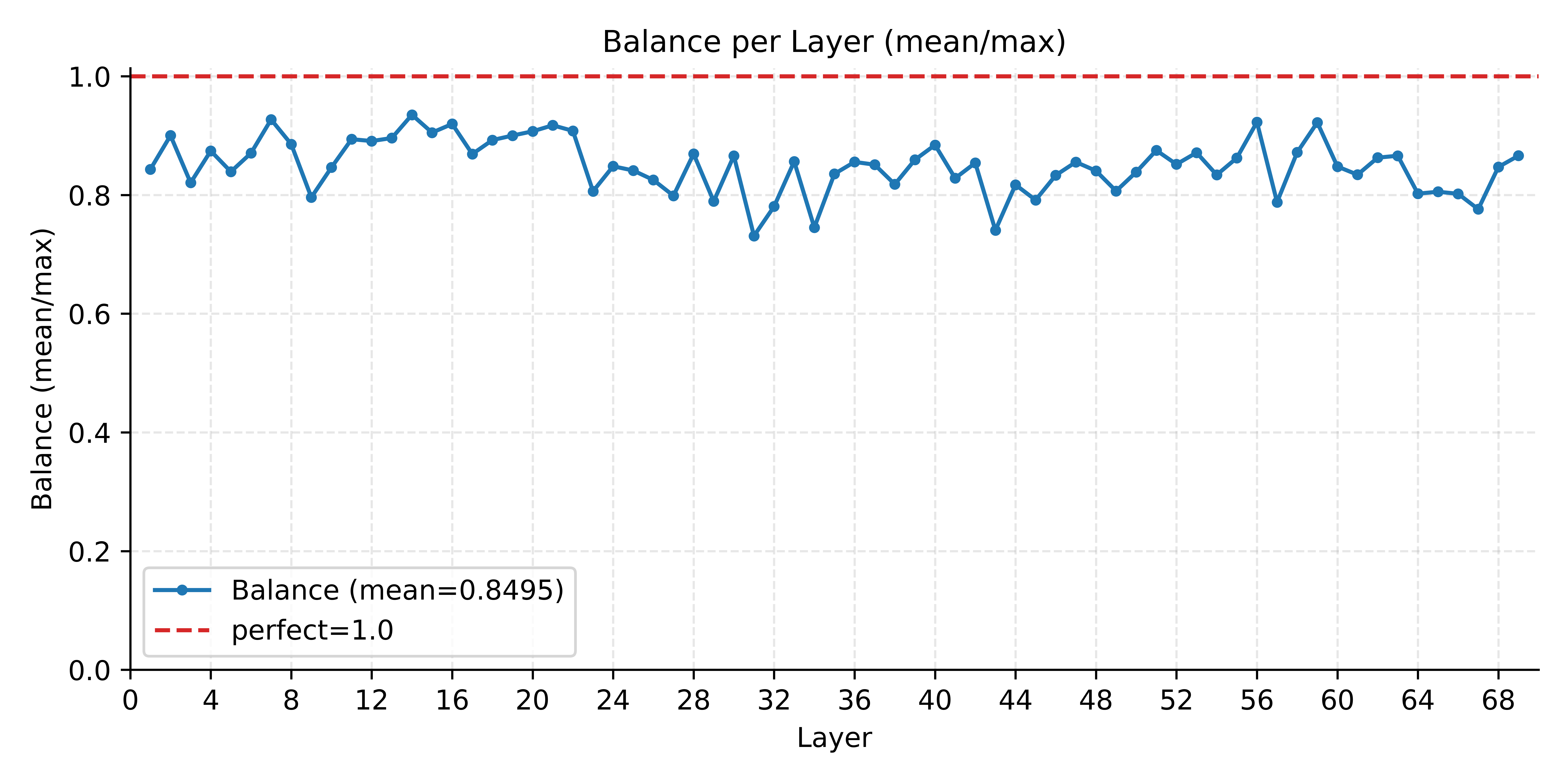}
\caption{Per-layer expert balance (mean/max token count ratio) across all layers, averaging 0.8495, close to the perfect value of 1.0.}
\label{fig:expert_balance}
\end{figure}

\subsection{Resolving NUMA Conflicts}

The \texttt{numa\_balancing} kernel parameter in certain Ubuntu systems conflicts with SGLang's numa-node configuration, causing sporadic large execution gaps between compute kernels during model inference. In multi-node multi-GPU deployments, these gaps appear at random positions across ranks, and each inter-rank synchronization is bottlenecked by the slowest rank --- significantly impacting overall inference efficiency. Disabling the system kernel's \texttt{numa\_balancing} parameter resolved the issue, \textbf{improving end-to-end performance by approximately 10\%}.

\section{Decode Optimization}

\subsection{GPU Memory Optimization}

In agentic scenarios, multi-turn conversations cause the context to grow continuously, making KVCache GPU memory usage the primary decode bottleneck --- once memory is filled by KVCache, batch size cannot expand, GPU compute units are not saturated, and decode throughput is limited, requiring more nodes to maintain throughput and driving up inference costs. To increase single-node concurrency, we implemented multiple memory optimizations:

\begin{enumerate}
    \item \textbf{Decode KVCache SWA support:} KVCache effective capacity increased to \textbf{${\sim}5\times$}.
    \item \textbf{PD-disaggregated KVCache preallocation optimization:} Moved the preallocation of KVCache for incoming requests from GPU memory to CPU memory, only transferring to GPU memory when decode actually starts, eliminating waste from resource over-provisioning.
    \item \textbf{CUDA Graph memory tuning:} Optimized CUDA Graph parameters to reduce wasted memory, increasing KVCache capacity.
\end{enumerate}

\subsection{MTP Optimization}

The MiMo-V2.5 series natively supports 3-layer MTP (Multi-Token Prediction) to accelerate decode output, but prefill previously did not enable MTP --- causing the first 128 decode output tokens to have invalid KVCache in the MTP layers, with very low prediction acceptance rates. Since agentic scenarios involve mostly short output sequences, this limitation significantly limited MTP's effective speedup. By introducing MTP support during prefill with dedicated adaptations and optimizations for HiCache L2/L3, MTP acceleration during the early decode phase improved substantially: \textbf{0--128 token speedup reached $2.3\times$, 128--256 token speedup reached $1.5\times$}, effectively reducing actual decode cost in agentic scenarios.

\section{Multimodal Inference Optimization}

Based on the SGLang community v0.5.7 EPD design, we performed a range of engineering optimizations and stability fixes for EPD disaggregation in the MiMo-V2.5 series, \textbf{doubling Encoder throughput} with no latency regression. We are upstreaming these changes to SGLang (issue \#24945). The Encoder performance before and after optimization is summarized in Table~\ref{tab:multimodal_perf}.

\begin{table}[t]
\centering
\caption{Encoder performance before and after optimization.}
\label{tab:multimodal_perf}
\begin{tabular}{lccc}
\toprule
 & QPS & Avg Latency (ms) & P90 Latency (ms) \\
\midrule
Before & 15 & 78.39 & 100.76 \\
After  & 30 & 80.28 & 82.94 \\
\bottomrule
\end{tabular}
\end{table}

\subsection{Architecture Optimization}

\begin{itemize}
    \item \textbf{Overlap multimodal embedding transfer with inference:} In the prefill scheduler's main loop, we support asynchronous replication of multimodal embedding data across TP ranks, overlapping it with prefill inference to reduce GPU idle time.
    \item \textbf{Data parallelism for the Encoder:} Since the Encoder model is relatively small, setting TP$>$1 degrades performance. We deploy Encoder with TP$=$1 while supporting data parallelism, simplifying single-machine 8-GPU deployment and operations.
    \item \textbf{Encoder cross-request batch support:} We introduced cross-request batching for the EPD Encoder Server. The Encoder scheduler aggregates concurrent requests by modality, merging multiple requests' image/audio into a single forward pass then splitting and returning results per request, addressing the low GPU utilization caused by per-request encoding.
\end{itemize}

\subsection{Preprocessing Optimization}

\begin{itemize}
    \item \textbf{GPU image preprocessing:} For large images, executing resize/normalize/patchify on CPU significantly increases end-to-end latency, so we ported preprocessing to GPU, eliminating the CPU bottleneck.
    \item \textbf{Parallel image download and decode:} We use multi-process downloading and PIL decoding, avoiding delays from serial download and GIL contention.
    \item \textbf{Multimodal download and forward parallelism:} In the initial Encoder implementation, data download and inference were serial both across and within batches, leaving the GPU idle during downloads. We decoupled download from inference with a message queue, overlapping download and inference within a batch.
    \item \textbf{Parallel video decoding:} We evenly split frame extraction indices into $N$ chunks, spawning an independent VideoDecoder per chunk and decoding them in parallel threads, reducing end-to-end Encoder latency for a 1-hour video \textbf{from 156~s to 23~s}.
\end{itemize}

\subsection{Cache Optimization}

\begin{itemize}
    \item \textbf{Encoder consistent hashing:} In multi-Encoder scenarios, Prefill round-robin Encoder selection reduces multimodal cache hit rates. Through consistent hashing, we route requests with the same key to the same Encoder, \textbf{improving cache hit rate by 30\%}.
    \item \textbf{Intra-node Embedding cache sharing:} Using shared memory, we enable multimodal cache data sharing across multiple Encoder GPUs on the same node, improving cache hit rate.
\end{itemize}

\section{Afterword}

Looking back, the inference efficiency of the MiMo-V2.5 series did not come from a single breakthrough, but from coordinated optimization across multiple dimensions. Hybrid SWA benefits both prefill and decode, but an insufficiently optimized KVCache implementation can actually increase costs in both stages. To address this, we systematically refactored KVCache management, tiered caching, and prefix cache trees, tackled the core challenges of SWA-aware KVCache, and optimized scheduling and the Prefill/Decode pipeline. All changes were validated in production, ultimately realizing Hybrid SWA's theoretical efficiency gains. Only then did Hybrid SWA fully realize its architectural advantage of combined performance and efficiency in long-context inference. Further optimizations to the MoE configuration and multimodal inference pipeline also substantially boosted serving performance.

We present the first large-scale engineering implementation that comprehensively covers the Hybrid SWA + MoE + multimodal composite architecture, and pass the resulting cost savings back to users through API price reductions. At the same time, we have contributed a subset of our optimizations to the SGLang open-source community via PRs and will continue advancing more open-source initiatives --- with the goal of making engineering optimization less of a barrier, so that these high-performance, high-efficiency composite architectures can be more broadly explored and adopted.

\appendix

\newpage
\bibliography{main}

@article{sglang,
  title={Sglang: Efficient execution of structured language model programs},
  author={Zheng, Lianmin and Yin, Liangsheng and Xie, Zhiqiang and Sun, Chuyue and Huang, Jeff and Yu, Cody H and Cao, Shiyi and Kozyrakis, Christos and Stoica, Ion and Gonzalez, Joseph E and others},
  journal={Advances in neural information processing systems},
  volume={37},
  pages={62557--62583},
  year={2024}
}

@misc{mimov25pro,
  author       = {{Xiaomi MiMo Team}},
  title        = {{MiMo-V2.5-Pro}},
  year         = {2026},
  howpublished = {\url{https://huggingface.co/XiaomiMiMo/MiMo-V2.5-Pro}}
}

@misc{mimov25,
  author       = {{Xiaomi MiMo Team}},
  title        = {{MiMo-V2.5}},
  year         = {2026},
  howpublished = {\url{https://huggingface.co/XiaomiMiMo/MiMo-V2.5}}
}

@misc{deepseekv4pro,
  author       = {{DeepSeek-AI}},
  title        = {{DeepSeek-V4-Pro}},
  year         = {2026},
  howpublished = {\url{https://huggingface.co/deepseek-ai/DeepSeek-V4-Pro}}
}

@misc{deepseekv4flash,
  author       = {{DeepSeek-AI}},
  title        = {{DeepSeek-V4-Flash}},
  year         = {2026},
  howpublished = {\url{https://huggingface.co/deepseek-ai/DeepSeek-V4-Flash}}
}

@misc{glm5,
  author       = {{Z.ai}},
  title        = {{GLM-5}},
  year         = {2026},
  howpublished = {\url{https://huggingface.co/zai-org/GLM-5}}
}

@misc{kimik2,
  author       = {{Moonshot AI}},
  title        = {{Kimi-K2-Instruct}},
  year         = {2025},
  howpublished = {\url{https://huggingface.co/moonshotai/Kimi-K2-Instruct}}
}

@misc{qwen35,
  author       = {{Qwen}},
  title        = {{Qwen3.5-397B-A17B}},
  year         = {2026},
  howpublished = {\url{https://huggingface.co/Qwen/Qwen3.5-397B-A17B}}
}

@misc{minimaxm2,
  author       = {{MiniMax}},
  title        = {{MiniMax-M2}},
  year         = {2025},
  howpublished = {\url{https://huggingface.co/MiniMaxAI/MiniMax-M2}}
}

@misc{hy3,
  author       = {{Tencent Hy}},
  title        = {{Hy3}},
  year         = {2026},
  howpublished = {\url{https://huggingface.co/tencent/Hy3}}
}
\newpage

\section{Contributions and Acknowledgments}
We would like to express our sincere gratitude to all contributors for their invaluable support and
efforts.
\textit{Authors within each role are listed alphabetically by their first name}.

{\renewcommand{\thefootnote}{\fnsymbol{footnote}}\footnotetext[2]{Corresponding author}}

\begin{multicols}{2} %
\noindent
Anqi Liu \\
Aoxin Ma \\
Bo Chen \\
Bo Yang \\
Chen Wang \\
Chen Zhang \\
Chengda Tang \\
Chengwei Wang \\
Chiheng Lou \\
Depeng Yan \\
Fuli Luo$^\dagger$ \\
Gang Wang \\
Hailin Zhang \\
Jiale Sun \\
Kang Zhou \\

Rui Huang \\
Shaohui Liu \\
Shen Huang \\
Shijie Cao \\  
Shuaishuai Fan \\
Tianling Zhou \\
Xiangwei Deng \\
Xueyang Xie \\
Xuli Wang \\
Yingchun Lai \\
Yu Yang \\
Yuan Zhang \\
Zhen Tang \\
Zhonghua Deng \\
Zihan Jiang \\

\end{multicols} %

\newpage

\end{document}